\documentclass[]{spie}  %>>> use for US letter paper
\usepackage{amsmath,amsfonts,amssymb}
\usepackage{graphicx}
\newcommand*\aap{A\&A}

\newcommand*\aj{AJ}

\newcommand*\apj{ApJ}
\newcommand*\apjl{ApJ}

\newcommand*\apjs{ApJS}

\newcommand*\mnras{MNRAS}

\newcommand*\nat{Nature}

\newcommand*\pasp{PASP}

\newcommand*\procspie{Proc SPIE}

\title{Hi-5: a potential high-contrast thermal near-infrared imager for the VLTI}
\authorinfo{Denis Defr\`ere: ddefrere@uliege.be, +323669758}

\author[a]{D. Defr\`ere}
\author[b]{M.~Ireland}
\author[a]{O. Absil}
\author[c]{J.-P.~Berger}
\author[d]{W.C.~Danchi}
\author[e]{S.~Ertel}
\author[f]{A.~Gallenne}
\author[c]{F.~H\'enault}
\author[e]{P.~Hinz}
\author[g]{E.~Huby}
\author[h]{S.~Kraus}
\author[i]{L.~Labadie}
\author[c]{J.-B.~Le Bouquin}
\author[c]{G.~Martin}
\author[j]{A.~Matter}
\author[k]{B.~Mennesson}
\author[l]{A.~M\'erand}
\author[m,p]{S.~Minardi}
\author[n]{J.D.~Monnier}
\author[o]{B.~Norris}
\author[a]{G.~Orban de Xivry}
\author[m,p]{E.~Pedretti}
\author[q]{J.-U.~Pott}
\author[a]{M.~Reggiani}
\author[k]{E.~Serabyn}
\author[a]{J.~Surdej}
\author[f]{K.\,R.\,W.~Tristram}
\author[l]{J.~Woillez}

\affil[a]{Space sciences, Technologies \& Astrophysics Research (STAR) Institute, University of Li\`ege, Li\`ege, Belgium}
\affil[b]{Research School of Astronomy and Astrophysics, Australian National University, Canberra, ACT 2611, Australia}
\affil[c]{University Grenoble Alpes, CNRS, IPAG, 38000 Grenoble, France}
\affil[d]{NASA Goddard Space Flight Center, Exoplanets \& Stellar Astrophysics Laboratory, Greenbelt, USA}
\affil[e]{Steward Observatory, Department of Astronomy, University of Arizona, Tucson, Arizona, USA}
\affil[f]{European Southern Observatory, Alonso de C\'ordova 3107, Vitacura, Santiago de Chile, Chile}
\affil[g]{LESIA, Observatoire de Paris, PSL Research University, 92195 Meudon Cedex, France}
\affil[h]{School of Physics and Astronomy, University of Exeter, Exeter, United Kingdom}
\affil[i]{I. Physikalisches Institut, Universit\"at zu K\"oln, Z\"ulpicher Str. 77, 50937 Cologne, Germany}
\affil[j]{Universit\'e C\^ote d'Azur, Observatoire de la C\^ote d'Azur, CNRS, Laboratoire Lagrange, Bd de l'Observatoire, CS 34229, 06304 Nice cedex 4, France}
\affil[k]{Jet Propulsion Laboratory, California Institute of Technology, Pasadena, CA 91109, USA}
\affil[l]{European Southern Observatory, Munich, Germany}
\affil[m]{University of Jena, Jena, Germany}
\affil[n]{University of Michigan, Ann Arbor, United States}
\affil[o]{University of Sydney, Sydney, Australia}
\affil[p]{innoFSPEC, Leibniz-Institut f\"{u}r Astrophysik Potsdam (AIP) Germany}
\affil[q]{Max Planck Institute for Astronomy, Heidelberg, Germany}

\begin{document} 
\maketitle

\begin{abstract}
Hi-5 is a high-contrast (or high dynamic range) infrared imager project for the VLTI. Its main goal is to characterize young extra-solar planetary systems and exozodiacal dust around southern main-sequence stars. In this paper, we present an update of the project and key technology pathways to improve the contrast achieved by the VLTI. In particular, we discuss the possibility to use integrated optics, proven in the near-infrared, in the thermal near-infrared (L and M bands, 3-5~$\mu$m) and advanced fringe tracking strategies. We also address the strong exoplanet science case (young exoplanets, planet formation, and exozodiacal disks) offered by this wavelength regime as well as other possible science cases such as stellar physics (fundamental parameters and multiplicity) and extragalactic astrophysics (active galactic nuclei and fundamental constants). Synergies and scientific preparation for other potential future instruments such as the Planet Formation Imager are also briefly discussed.
\end{abstract}

\section{Introduction}
\label{intro}

\begin{figure}[!b]
	\begin{center}
		\includegraphics[height=11.2 cm]{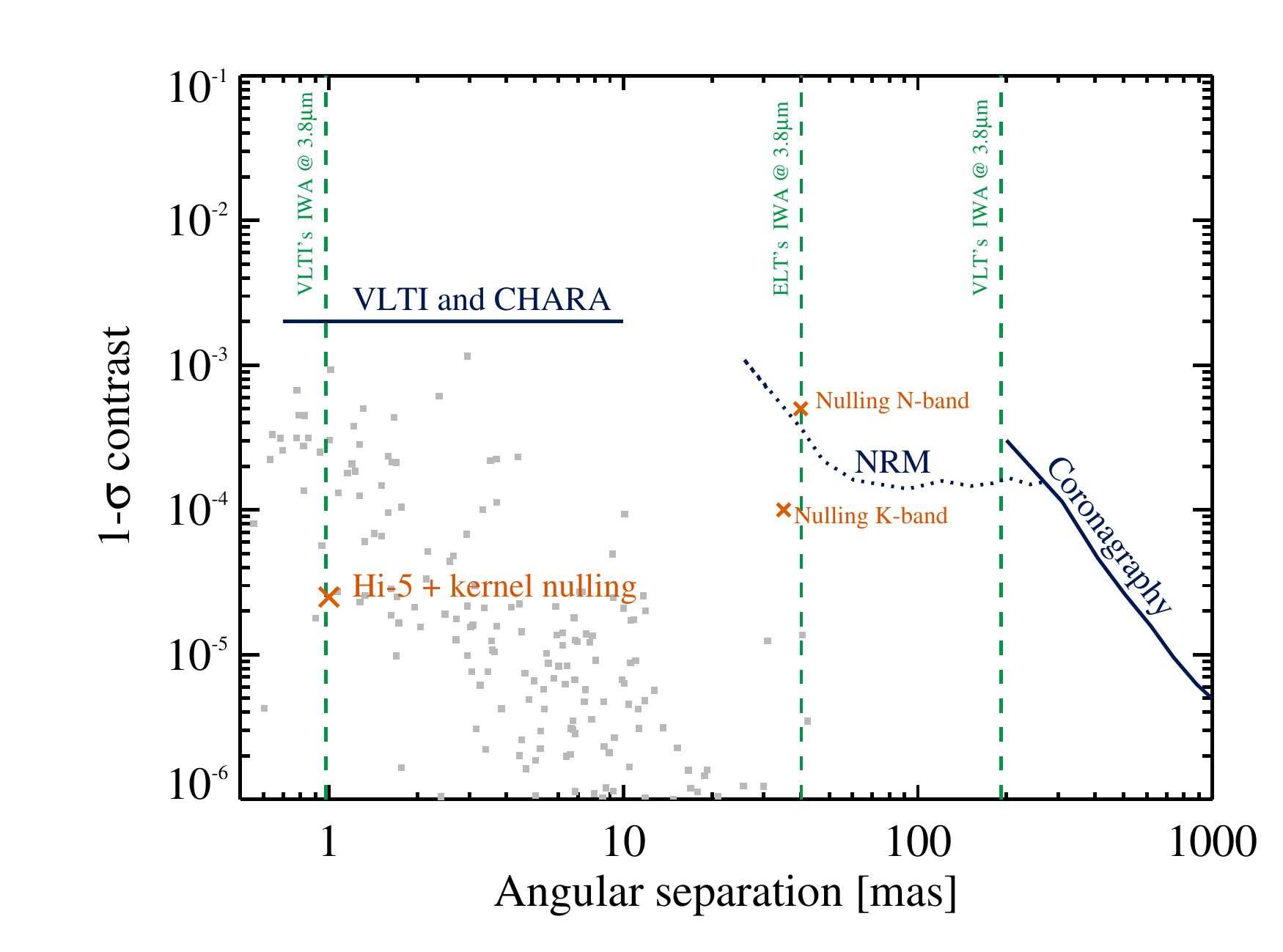}
		\caption{Contrast level as a function of angular separation for current CHARA and VLTI instruments (upper left horizontal blue line) compared to that of nulling interferometers installed in the Northern hemisphere (orange crosses, PFN for K-band\cite{Hanot:2011} and LBTI for N-band\cite{Defrere:2016}), non-redundant aperture masking (NRM) experiments operating at L-band\cite{Hinkley:2011} (dotted line), and single-dish L-band coronagraphs\cite{Absil:2016} (righ solid blue line). The expected performance of Hi-5 is indicated by the lower left orange cross. From left to right, the vertical dashed lines represent the inner working angle at L band of the VLTI, the ELT, and the VLT (computed as $0.25\times\lambda/b$ for the interferometers and as $2\times\lambda/D$ for the single-aperture instruments). Known exoplanets discovered by radial velocity surveys are over-plotted as gray squares.}\label{fig1}
		\vspace{-1.5em}
	\end{center}
\end{figure}

Hi-5 (High-contrast Interferometry up to 5\,$\mu$m) is a project for a new high-contrast instrument for the VLTI. While similar projects were already discussed in the past\cite{Absil:2006}, this idea came back recently thanks to recent developments in integrated optics for the thermal near-infrared (L and M bands, 3-5~$\mu$m) as well as the demonstration of record-breaking high-contrast nulling interferometry both in the near-infrared (K band) on the Palomar Fiber Nuller (PFN \cite{Mennesson:2011}) and in the mid-infrared (N band) on the Large Binocular Telescope Interferometer (LBTI \cite{Defrere:2016}, see Figure~\ref{fig1}). Hi-5 was first discussed at the SPIE meeting in Montr\'eal in 2014 and was selected a couple of years later by the European Interferometry Initiative (EII) as one of the two most promising concepts for a future VLTI instrument. It is also now discussed as a possible future instrument in the VLTI roadmap\cite{Merand:2018} and received funds from the H2020 OPTICON Joint Research Network as part of a larger interferometry proposal. It also received funds from the University of Li\`ege to organize a kickoff meeting in Li\`ege.  This meeting took place in October 2017 and gathered $\sim$25 experts in the fields of high-contrast imaging, interferometry, integrated optics, fringe tracking, and interferometry science cases (see http://www.biosignatures.ulg.ac.be/hi-5/index.html). The discussions and presentations of this meeting were summarized and recently published in Experimental Astronomy\cite{Defrere:2018b}. In this paper, we give an update on the project since the kick-off meeting. We present recent progress in both the preparation of the science case (Section~\ref{sec:2}) and the technology roadmap (Section~\ref{sec:3}). Finally, we discuss in Section~\ref{sec:4} the synergies and scientific preparation for other potential future major interferometric instruments or facilities.

\section{Hi-5 science cases}
\label{sec:2}

\subsection{Characterization of known exoplanets}

Nearly 4000 confirmed exoplanets have been discovered so far but only about 45 planets or low-mass brown dwarfs have been directly observed\cite{Bowler:2016}. Directly detecting the photons from the planetary photosphere provides a wealth of information on the nature of the observed exoplanet but is also very challenging due to the small angular resolution and the huge contrast between a planet and its host star. Assuming an IWA of 1\,mas (0.25*$\lambda$/B), an outer working angle of 250\,mas, and a contrast of at least 10$^{-5}$, the VLTI would be able to characterize approximately 30 known exoplanets discovered by radial velocity as shown in Table~\ref{tab_targets}. All but one of these planets are located within 10\,mas from their host star and most of them at a few mas (see also Figure~\ref{fig1}), which would require the long VLTI baselines in order to be observable. Low-resolution spectroscopic observations of such planets in the thermal near-infrared will provide their radius and effective temperature as well as critical information to study the non-equilibrium chemistry of their atmosphere via the CH$_4$ and CO spectral features. This could potentially lead to constraints on the composition and formation process of these exoplanets\cite{Konopacky:2013,Barman:2015,Skemer:2016}. 
\vspace{-0.2cm}
\begin{table}[!b]  
\caption{List of known exoplanets and brown dwarfs located within 50\,pc that can be observed from Paranal and could be observed with the VLTI at L-band (assuming an IWA of 1\,mas, an outer working angle of 250\,mas, and a contrast of at least 10$^{-5}$). All these exoplanets will be inaccessible with the ELT as shown in Figure~\ref{fig1}. The target list is built from the exoplanet catalog (http://exoplanet.eu/).}\label{tab_targets}
\begin{center}    
\begin{tabular}{ c | c c c c c | c c c c c} 
\hline
\hline
	 & \multicolumn{5}{c}{Star} & \multicolumn{5}{|c}{Planet}\\
\hline
Name & Dist. & RA    & DEC   & V$_{\rm mag}$ & K$_{\rm mag}$ & Mass & T$_{\rm eff}$ & Sep. &  Sep. & Contrast\\
	 & [pc]  & [deg] & [deg] &      			&     		    & [M$_{\rm Jup}$] & [K]  & [mas] & [AU] & log$_{10}$\\
\hline
tau Boo b & 15.6 & 206.8 & 17.5 & 4.5 & 3.4 & 5.84 & 1637 & 3 & 0.046 & -2.94\\
HD 162020 b & 31.3 & 267.7 & -40.3 & 9.2 & 6.5 & 14.40 & 744 & 2 & 0.074 & -3.21\\
HD 179949 b & 27.0 & 288.9 & -24.2 & 6.3 & 4.9 & 0.92 & 1629 & 2 & 0.045 & -3.36\\
HD 75289 b & 28.9 & 131.9 & -41.7 & 6.4 & 5.1 & 0.47 & 1322 & 2 & 0.046 & -3.63\\
HD 217107 b & 19.7 & 344.6 & -2.4 & 6.2 & 4.5 & 1.33 & 1019 & 4 & 0.073 & -3.65\\
HD 102195 b & 29.0 & 176.4 & 2.8 & 8.1 & 6.1 & 0.45 & 1129 & 2 & 0.049 & -3.67\\
HD 77338 b & 40.8 & 135.3 & -25.5 & 8.6 & 6.7 & 0.50 & 1013 & 2 & 0.061 & -3.83\\
HD 195019 b & 37.4 & 307.1 & 18.8 & 6.9 & 5.3 & 3.70 & 768 & 4 & 0.139 & -3.95\\
HD 285507 b & 41.3 & 61.8 & 15.3 & 10.5 & 7.7 & 0.92 & 734 & 2 & 0.073 & -3.98\\
HD 38529 b & 39.3 & 86.6 & 1.2 & 5.9 & 4.2 & 0.93 & 1104 & 3 & 0.131 & -4.00\\
%HD 159243 b & 69.2 & 263.3 & 5.7 & 8.7 &  & 1.13 & 916 & 2 & 0.110 & -4.02\\
GJ 86 A b & 10.9 & 32.6 & -50.8 & 6.2 & 4.1 & 4.01 & 651 & 10 & 0.110 & -4.03\\
HD 130322 b & 30.0 & 221.9 & 0.3 & 8.1 & 6.2 & 1.05 & 719 & 3 & 0.088 & -4.17\\
HD 168746 b & 43.1 & 275.5 & -11.9 & 8.0 & 6.2 & 0.23 & 932 & 2 & 0.065 & -4.18\\
HD 108147 b & 38.6 & 186.4 & -64.0 & 7.0 & 5.7 & 0.26 & 1006 & 3 & 0.102 & -4.32\\
BD-06 1339 b & 20.0 & 88.3 & -6.0 & 6.7 & 6.3 & 0.03 & 914 & 2 & 0.043 & -4.48\\
%HD 72892 b & 72.8 & 128.7 & -14.5 & 8.8 &  & 5.45 & 577 & 3 & 0.228 & -4.53\\
HD 10180 c & 39.4 & 24.5 & -60.5 & 7.3 & 5.9 & 0.04 & 1130 & 2 & 0.064 & -4.55\\
%BD+20 2457 b & 200.0 & 154.2 & 19.9 & 9.8 &  & 21.42 & 628 & 7 & 1.450 & -4.56\\
%HD 110014 c & 90.0 & 189.8 & -8.0 & 4.7 &  & 3.10 & 732 & 7 & 0.640 & -4.57\\
%11 Com b & 110.6 & 185.2 & 17.8 & 4.7 &  & 19.40 & 642 & 12 & 1.290 & -4.59\\
%HD 102272 b & 360.0 & 176.6 & 14.1 & 8.7 & 6.26 & 5.90 & 655 & 2 & 0.614 & -4.60\\
%ksi Aql b & 62.7 & 298.6 & 8.5 & 4.7 & 2.17 & 2.02 & 780 & 9 & 0.580 & -4.61\\
GJ 876 d & 4.7 & 343.3 & -14.3 & 10.2 & 5.0 & 0.02 & 625 & 4 & 0.021 & -4.65\\
HD 27894 b & 42.4 & 65.2 & -59.4 & 9.4 &  & 0.67 & 611 & 3 & 0.125 & -4.66\\
61 Vir b & 8.5 & 199.6 & -18.3 & 4.7 & 3.0 & 0.02 & 1145 & 6 & 0.050 & -4.67\\
%HD 33283 b & 86.0 & 77.0 & -26.8 & 8.1 &  & 0.33 & 817 & 2 & 0.168 & -4.67\\
%HIP 105854 b & 80.8 & 321.6 & -37.8 & 5.6 &  & 8.20 & 630 & 10 & 0.810 & -4.68\\
%HD 224693 b & 94.0 & 360.0 & -22.4 & 8.2 &  & 0.71 & 744 & 2 & 0.233 & -4.72\\
GJ 3998 b & 17.8 & 259.0 & 11.1 & 10.8 & 6.8 & 0.01 & 793 & 2 & 0.029 & -4.73\\
%HIP 107773 b & 106.0 & 327.5 & -64.7 & 5.6 &  & 1.98 & 770 & 7 & 0.720 & -4.75\\
HD 1461 b & 23.4 & 4.7 & -8.1 & 6.6 & 4.9 & 0.02 & 1094 & 3 & 0.063 & -4.76\\
%HD 179079 b & 63.7 & 287.8 & -2.6 & 8.0 &  & 0.08 & 885 & 2 & 0.110 & -4.78\\
HD 40307 b & 12.8 & 88.5 & -60.0 & 7.2 & 4.8 & 0.01 & 954 & 4 & 0.048 & -4.81\\
%HD 74156 b & 64.6 & 130.6 & 4.6 & 7.6 &  & 1.78 & 620 & 5 & 0.292 & -4.84\\
mu Ara c & 15.3 & 266.0 & -51.8 & 5.2 & 3.7 & 0.03 & 967 & 6 & 0.091 & -4.86\\
nu Oph b & 46.8 & 269.8 & -9.8 & 3.3 & 1.1 & 24.00 & 595 & 41 & 1.900 & -4.86\\
HD 168443 b & 37.4 & 275.0 & -9.6 & 6.9 & 5.2 & 7.66 & 496 & 8 & 0.293 & -4.87\\
%HIP 67851 b & 66.0 & 208.5 & -35.3 & 6.2 &  & 1.38 & 649 & 7 & 0.460 & -4.91\\
HD 192263 b & 19.9 & 303.5 & -0.9 & 7.8 &  & 0.73 & 559 & 8 & 0.153 & -4.92\\
HD 69830 b & 12.6 & 124.6 & -12.6 & 6.0 & 4.2 & 0.14 & 829 & 6 & 0.079 & -4.94\\
HD 102117 b & 42.0 & 176.2 & -58.7 & 7.5 & 5.8 & 0.17 & 711 & 4 & 0.153 & -4.98\\
HD 16417 b & 25.5 & 39.2 & -34.6 & 5.8 & 4.2 & 0.07 & 851 & 5 & 0.140 & -4.99\\
\hline
\end{tabular}
\end{center}
\end{table}

\subsection{Dedicated survey of young stellar clusters}

\begin{figure}[b!]
      \centering
      \includegraphics[scale=0.5]{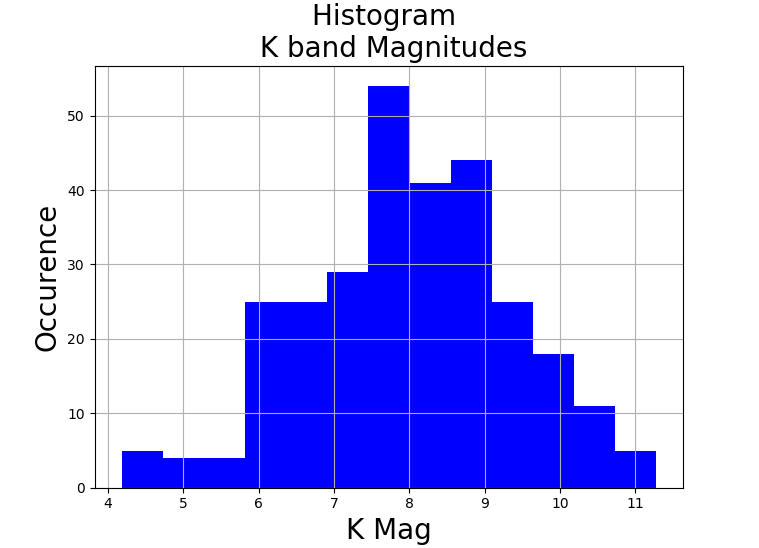}
      \caption{Brightness distribution of all young stars presented in Table 1.}
      \label{fig3}
\end{figure}

The scarcity of giant exoplanets discovered by ongoing single-aperture direct imaging surveys currently challenges our understanding and theories of planet formation. For the nearest population of young stars, surveys are typically sensitive to planets further than 10~AUs from their host stars due to the limited angular resolution. This gives access to the semi-major axes where giant planets could have formed in-situ\cite{Ormel:2010,Lambrechts:2012} but does not fill the gap with the exoplanet population discovered by radial velocity surveys. Filling this gap is very important because it can provide critical constraints on planet formation theories and evolution models \cite{Spiegel:2012,Mordasini:2012,Allard:2013}. With a new high-contrast VLTI instrument, it will be possible to further fill this parameter space by resolving exoplanets closer to their host star. Focusing on young stars, the thermal near-infrared provides an optimum sensitivity to cooler planets, including lower-mass planets, and planets that are born cold because they accrete their envelopes through a radiatively efficient shock\cite{Marley:2007}. A survey of nearby young stellar moving groups could detect new giant planets at angular distances inaccessible by current instruments and future ELTs. As shown in Table~\ref{tab1} and Figure~\ref{fig3}, approximately 280 young ($<$50~Myr) and relatively bright (K$<$10) stars can be observed from Paranal. Monte-Carlo simulations are currently being performed to estimate the number of new detections that can be expected for different performance levels (contrast, sensitivity, and brightness of the targets).

\begin{table}[!b]
\caption{List of nearby moving groups\cite{Bell:2015} accessible from Paranal and given with their age and minimum planetary mass ($M_{min}$, in Jupiter masses) that can be detected assuming a contrast performance of 10$^{-4}$. $N_{tot}$ is the total number of stars in the group. $N_{J \le 10mag}$ and $N_{K \le 10mag}$ are the number of targets brighter than J=10 and K=10 respectively.} 
\label{tab1}
\begin{center}       
\begin{tabular}{ c c c c c c} 
\hline
\hline
\rule[-1ex]{0pt}{3.5ex} Name & Age(Myr) & $M_{min}(M_{\rm Jup})$ & $N_{tot}$ & $N_{J \le 10mag}$& $N_{K \le 10mag}$\\
\hline
%\rule[-1ex]{0pt}{3.5ex} AB Dor & $149^{+51}_{-19}$ & $\varnothing$ & 89 & 52 & 58  \\
%\rule[-1ex]{0pt}{3.5ex} Argus & $69^{+19}_{-8}$ & $\varnothing$ & 27 & 15 & 16 \\
\rule[-1ex]{0pt}{3.5ex} $\eta$ Cha & 11 $\pm$ 3 & 2.1 & 18 & 0 & 0\\
\rule[-1ex]{0pt}{3.5ex} Carina & $45^{+11}_{-7}$ & 3.1 & 12 & 4 & 5\\
\rule[-1ex]{0pt}{3.5ex} Columba & $42^{+6}_{-4}$ & 3.1 & 50 & 26 & 38\\
\rule[-1ex]{0pt}{3.5ex} TWA & 10$\pm$3 & 2.1 & 30 & 18 & 17\\
\rule[-1ex]{0pt}{3.5ex} Tuc-Hor & 45$\pm$4 & 4.2 & 189 & 80 & 149  \\
\rule[-1ex]{0pt}{3.5ex} BPMG & 24$\pm$3 & 2.1 & 97 & 60 & 61 \\
\rule[-1ex]{0pt}{3.5ex} 32 Ori & 22$\pm$4 & 2.1 & 14 & 8 & 11\\
\hline
\rule[-1ex]{0pt}{3.5ex} All & $\varnothing$ & $\varnothing$ & 410 & 196 & 281  \\
\hline
\end{tabular}
\end{center}
\end{table}

\subsection{Young stellar objects}

Planets form in the disks around young stars. During the first few million years, these disks are optically thick and the planetary cores are deeply embedded in the disk material. As the planets interact with the disk and the disk dissipates, the planets should become observable through direct imaging. Most planet searches with interferometry in young systems have been conducted using the non-redundant aperture masking (NRM) technique \cite{Kraus:2012}. However, it has been found that asymmetric emission from the optically thick circumstellar disk can introduce strong phase signals, which can lead to false companion detections \cite{Olofsson:2013,Willson:2016}. Hi-5 will mitigate these problems by using 10 to 20 times longer baselines than single-aperture NRM interferometry, allowing us to better separate the planet emission from the disk emission. Such observations will serve as important exploratory science for the Planet Formation Imager (PFI, also described in this volume). 

\subsection{Exozodiacal disks}\label{sec:exozodi}

Exozodiacal dust emits primarily in the near-infrared to mid-infrared where it is outshone by the host star. Due to the small angular scales involved (1\,AU at 10\,pc corresponds to 0.1\,arcsec), the angular resolution required to spatially disentangle the dust from the stellar emission currently requires the use of interferometry. Thus, exozodis have so far mostly been observed at the CHARA array and the VLTI in the near-infrared \cite{abs06, Absil:2013, def12b, Ertel:2014, ert16} and at KIN and the LBTI in the mid-infrared \cite{mil11, men13, def15, Ertel:2018}. These observations reached contrasts of a few 10$^{-4}$ to a few 10$^{-3}$, leading to vital statistical insights into the occurrence rates of exozodis as a function of other properties of the systems such as the presence of cold, Kuiper belt-like dust disks or stellar age and spectral type. The main challenge at the moment is linking the near-infrared and the mid-infrared detections, which critically constrain the systems' architectures and the properties and origin of the dust. However, so far no connection between the detections in the two wavelength ranges has been found. A high-contrast instrument operating in the thermal near-infrared like Hi-5 is the ideal tool to trace the spectral energy distributions of near-infrared detected exozodis toward longer wavelengths and of mid-infrared detected exozodis toward shorter wavelengths in order to connect the two and to understand non-detections in one wavelength range in the light of detections in the other. Moreover, no sensitive interferometric instrument operating in the thermal near-infrared is available in the Southern hemisphere so far. MATISSE is not designed for high-contrast observations and will be limited to the characterization of the brightest systems already detected in the near-infrared. The high contrast thermal near-infrared capabilities of Hi-5, together with the efficiency increase due to the simultaneous use of four telescopes already demonstrated at the VLTI with PIONIER, will allow for a large survey of habitable zone dust in the Southern hemisphere. Estimates performed during the GENIE study suggested that a contrast of a few 10$^{-4}$ will be enough to study habitable zone dust down to levels approximately 50 times as dense as the Solar zodiacal cloud \cite{Absil:2006}. This will significantly improve our understanding of the occurrence rates of systems harboring Solar-like  exozodiacal dust and increase the currently very short list of such systems that can be studied in detail. 

\subsection{Stellar physics: binarity accross the HR diagram}

Optical interferometry has been used to complement AO assisted imaging survey of stars to estimate the multiplicity fraction. For instance, it has been shown that virtually all massive stars are in multiple systems\cite{Sana:2014}, thanks to a distance-limited survey of close-by massive stars. Extending this result to other classes of stars is still to be done, and one of the limitations is contrast: companion detection requires both inner-working angle and detection depth. Apart from multiplicity fraction, another interest of binarity study is determination of the fundamental parameters such as dynamical masses and distances. Spectroscopy is currently more sensitive than interferometry, so the stellar mass is only known to the sine of the inclination of the orbit. Direct imaging and follow up of the companion allows to estimate the true stellar mass. For example, companions around Cepheid pulsating stars are difficult to detect and only the brightest companions are detected using near-infrared interferometry in the 10\,mas separation / 5$\times$10$^{-3}$ contrast regime \cite{Gallenne:2013,Gallenne:2014}. The perspective of spectroscopic radial velocities of the companion (using UV spectroscopy) and interferometric visual orbit opens the possibility for independent distance to Cepheids, rivaling Gaia in terms of distance accuracy (Gallenne et al., in preparation). Even if the L band is not optimum to detect hot Cepheid companions, a ten-fold improvement in contrast compared to current H-band instruments will still lead to more detections.

\subsection{Extragalactic astrophysics} 

Most of the physical processes in Active Galactic Nuclei (AGN) take place on scales of a few parsec or less (i.e.\ $\lesssim 100 \, \mathrm{mas}$ for the nearest galaxies). Hence, it requires interferometic methods to resolve the relevant scales. AGN are ``faint'' for infrared interferometry, with fluxes of $F_\mathrm{K} < 70 \, \mathrm{mJy}$ ($K > 10 \, \mathrm{mag}$) in the near-infrared, and $F_\mathrm{N} < \, 1\mathrm{Jy}$ ($N > 4 \, \mathrm{mag}$, with a few exceptions) in the mid-infrared. Additionally, AGN spectra are very red and they often appear extended in the optical, leading to limitations for fringe tracking and poor AO correction. Nevertheless, interferometric observations of several AGN with the VLTI and the Keck Interferometer have shown that their dust distributions are compact, with sizes roughly scaling with the square root of the intrinsic luminosity \cite{Tristram:2011, Kishimoto:2011, Burtscher:2013}. The few better resolved sources reveal a two component structure, with a central disk and an emission extending in the polar direction \cite{Hoenig:2012, Tristram:2014, Lopez-Gonzaga:2016}. However, the number of AGN observable by current instruments is very limited and most sources only appear marginally resolved, especially towards shorter wavelengths. Further progress in our understanding of AGN can hence only be expected from an instrument providing \emph{high accuracy visibility} (or high-contrast) measurements as well as a \emph{high sensitivity}. This will allow to better constrain larger samples of marginally resolved AGN, especially if also the ATs on the longest baselines can be used. Finally, of a different scientific interest, by combining interferometric with reverberation measurements, direct distances to such sources can be determined \cite{Hoenig:2014}, with the possibility to independently constrain the Hubble parameter.

\section{Improving the contrast of the VLTI}\label{sec:3}

\subsection{Instrumental architecture}

\begin{figure}[!t]
	\begin{center}
		\includegraphics[height=11.5 cm]{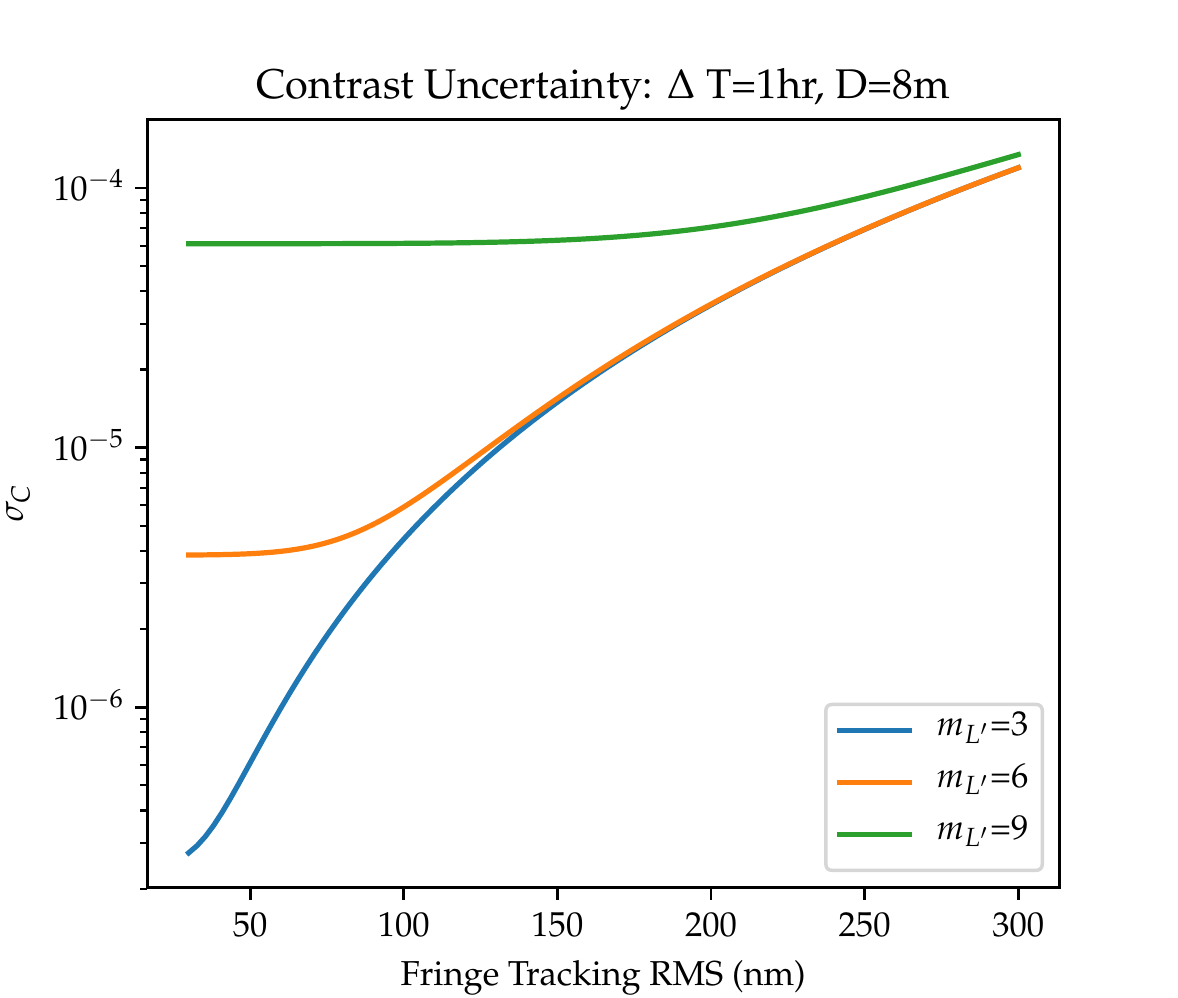}
		\caption{Contrast uncertainty as a function of fringe tracker performance for a L-band VLTI instrument using kernel nulling\cite{Martinache:2018}. Each line represents a different target L'-band magnitude. }\label{fig2}
		\vspace{-1.5em}
	\end{center}
\end{figure}

Over the past few years, several new interferometer architectures have been proposed in order to improve the performance and robustness against perturbations of nulling interferometers. In particular, an architecture concept that produces closure-phase measurements from nulled outputs has been proposed \cite{Lacour:2014}. More recently, the idea of ``kernel nulling'' has been proposed as a more generalized approach\cite{Martinache:2018}. The idea behind these concepts is to combine the outputs of a first nulling stage in a second mixing stage, with the goal of creating an output signal more robust against imperfect cophasing of the incoming stellar light. With such an architecture, contrasts of at least 10$^{-5}$ are in theory possible for reasonable fringe tracking performance of 150~nm RMS as shown in Figure~\ref{fig2}.

\subsection{Fringe tracking}

Phase-referenced interferometers require accurate and robust fringe tracking for sensitive background-limited observations and high-contrast imaging. Studies performed more than a decade ago concluded that closed-loop OPD residuals of no more than $\sim10$\,nm RMS are required to reach contrasts of a few 10$^{-4}$ at 3.8~$\mu$m \cite{Serabyn:2000, Absil:2006}. More recently, new pre-processing and post-processing techniques have emerged to significantly relax these constraints. A promising pre-processing technique currently under investigation uses a dual fringe tracking and low-order adaptive optics concept based on a combination of non-redundant aperture interferometry and eigen-phase in asymmetric pupil wavefront sensing \cite{Martinache:2016}. Applied to the VLTI with the NAOMI adaptive optics system functioning on the ATs, this approach could in theory enable L-band contrasts of a few 10$^{-6}$  for bright stars (m$_{\rm L'}\sim$3 and fringe tracking sensitivity down to H$\simeq$12 at the level of $\sim$300\,nm RMS in most seeing conditions\cite{Martinache:2018}. Currently, the GRAVITY fringe tracker achieves a limiting magnitude of K$\simeq$7.0 with the ATs and K$\simeq$10.0 with the UTs in single-field mode, with performance levels of 200-300\,nm RMS, depending on conditions. Under very good conditions, 1 to 2 magnitudes deeper can be achieved but generally at the expense of stability.\\

Regarding post-processing techniques, a new statistical data reduction method, called null self calibration (NSC), has recently shown that it is possible to reach contrasts down to $10^{-4}$ with significantly relaxed instrumental constraints \cite{Hanot:2011,Mennesson:2011}. For instance, a contrast of a few $10^{-4}$ has been achieved with the LBTI despite closed-loop OPD residuals of approximately 400\,nm RMS \cite{Defrere:2016,Mennesson:2016}. The idea of this data reduction technique is to use the full distribution of null measurements and fit models of null histograms to the observed data. The great advantage of this approach is to be immune to errors on the nulling setpoint, which is one of the major issues of ground-based nulling interferometers. Currently only applicable to two-telescope interferometers, this method has to be generalized for more telescopes in order to be used with the VLTI.\\
 
Another limiting factor of current high-contrast long-baseline interferometers is the phase chromaticism induced by random fluctuations in the water vapor differential column density above each aperture (or water vapor seeing). This component of the OPD is not correctly tracked at the wavelength of the science channel when this one operates at a wavelength different from that of the fringe sensor. The impact of this effect on infrared interferometry has been addressed extensively in the literature, either in a general context \cite{Colavita:2004} or applied to specific instruments that include phase-referenced modes using K-band light such as VLTI/MIDI \cite{Meisner:2003,Matter:2010,Pott:2012}, the KIN \cite{Colavita:2010b}, and the LBTI \cite{Defrere:2016b}. A possible implementation to mitigate this effect could be to have two separate fringe trackers: a fast one operating at short wavelength to correct for fast OPD variations and a slow one operating at the same wavelength as the science channel to correct for water vapor seeing. This effect will be seriously considered in the context of the Hi-5 study. 

\subsection{Integrated optics}

Integrated optics (IO) is a key component of current high-contrast VLTI interferometers such as PIONIER (H band) and GRAVITY (K band). Indeed, thanks  to  their  single-mode  properties, IO components deliver a much more stable instrumental transfer function than equivalent bulk-optics beam combiners. Significant efforts have been made over the past few years to make IO components at longer wavelengths and the technology has now reached a level of maturity sufficient for astronomical considerations. In the thermal near-infrared, recent studies have been targeting the development of components with ultrafast laser inscription in mid-infrared-transparent glasses. Ultrafast laser inscription (ULI) is a versatile technique using highly focused pulses from a femtosecond laser to induce permanent structural modifications in a large variety of glasses \cite{Gattass:2008}. The modifications are responsible for localised changes of the refractive index, which can be used to manufacture photonic devices based on waveguides. Remarkably, three dimensional structures can be written by scanning the glass samples under the laser focus. Particularly interesting for thermal near-infrared interferometry is the processing of chalcogenide glasses such as Gallium Lanthanum Sulfide \cite{Rodenas:2012} or Germanium Arsenic Sulfide \cite{DAmico:2014}, which have transparency windows extending to a wavelength of about 10~$\mu$m. Waveguides with  propagation losses at the 0.7-0.9 dB/cm level are routinely manufactured with ULI techniques. Photonic building blocks such as Y-junctions \cite{Rodenas:2012} and 2x2 directional couplers have been demonstrated \cite{Arriola:2014}. Couplers similar to the latter component were recently characterised in the L-, L$^{\prime}$- and M-bands, demonstrating high broadband contrasts, low spectral phase distortion, and 30\% to 60\% measured throughput \cite{Tepper:2017a,Tepper:2017b}. More advanced components, allowing the combination of several telescopes, have also been manufactured and tested. The first component was a 3-telescope N-band combiner based on cascaded Y-junctions with three-dimensional avoidance of waveguide cross-overs \cite{Rodenas:2012}. More recently, a 2-telescope ABCD combination unit \cite{Benisty:2009} and a 4-telescope beam combiner based on Discrete Beam Combiner geometry \cite{Minardi:2010} were manufactured with ULI and tested interferometrically with monochromatic light at 3.39~$\mu$m \cite{Diener:2017}. Both components showed that retrieval of complex visibilities with high signal to noise is possible at relatively low illumination levels.\\

An alternative to ultrafast laser inscription to fabricate integrated optics beam combiners is the use of classical methods, such as Ti:indiffusion inside electro-optic crystals. These waveguides are interesting as the refractive index of the material, and therefore the phase of the propagating optical beam, can be modified by the application of an external voltage. In the particular case of Lithium Niobate crystals, the transparency window reaches 5.2~$\mu$m allowing to cover L and M bands. Using this technology, phase and intensity modulators \cite{Heidmann:2012}, achieving on-chip fringe scanning, fringe locking and high-contrast interferometry have been demonstrated using a monochromatic laser at 3.39~$\mu$m \cite{Martin:2014a}. Concepts such as active 2-Telescope ABCD \cite{Heidmann:2011} and 3-Telescope AC \cite{Martin:2014b} infrared beam combiners have been validated experimentally. However, propagation losses in these systems are currently too high (typically 5 dB/cm). Therefore, novel methods such as ULI presented above, but developed in electro-optic crystals, are being tested for waveguide fabrication, showing low propagation losses (1.5 dB/cm) in the first prototypes \cite{Nguyen:2017}. Finally, note that two-dimensional photolithography on a platform with Ge,As,Se and Ge,As,S based glasses offsets the potential for less than 0.5dB/cm losses in mm-scale chips tolerant to low bend-radius \cite{Kenchington:2017}. Comparing these different technologies and how their technical specifications translate into the usual interferometric metrics (e.g., sensitivity, limiting magnitude, instrumental contrast) will be one of the major goals of the Hi-5 system study.

\section{Synergies with other instruments}\label{sec:4}

Hi-5 will be complementary to several future high-angular resolution instruments operating in the thermal near-infrared as described below.

\begin{itemize}

\item MATISSE, Multi AperTure mid-Infrared SpectroScopic Experiment \cite{Lopez:2014, Matter:2016a}, is the second-generation thermal near-infrared and mid-infrared spectrograph and imager for the VLTI. MATISSE will provide a wide wavelength coverage, from 2.8 to 13~$\mu$m, associated with a milli-arcsecond scale angular resolution (3 mas in L band; 10 mas in N band), and various spectral resolutions from R$\sim$30 to R$\sim$5000. In terms of performance, many instrumental visibilities were measured in laboratory conditions during the test phase. Those measurements were performed with a very bright artificial IR source to estimate the instrumental contribution to the accuracy without being limited by the fundamental noises. Such a source would have an equivalent flux, if observed with the UTs, of 20 to 70 Jy in N-band, 400 Jy in M-band, and 600 Jy in L band. Computed over 4 hours in LM band and 3 days in N band, the absolute visibility accuracy is better than 0.5 percent in L band, 0.4 percent in M band, and 2.5 percent in N band, on average over the corresponding spectral band. Those promising results are extensively described in internal ESO documents written by the MATISSE consortium (private communication with A.~Matter). Eventually, the on-sky tests (commissioning), starting in March 2018, will provide the real on-sky performance (sensitivity, accuracy) of MATISSE, which will include the effects of the sky thermal background fluctuations, the atmospheric turbulence, and the on-sky calibration.

\item ELT/METIS \cite{Brandl:2016} is the Mid-infrared E-ELT Imager and Spectrograph for the European Extremely Large Telescope. METIS will provide diffraction limited imaging and medium resolution slit spectroscopy in both the thermal near-infrared and the mid-infrared (5-19~$\mu$m) ranges, as well as high resolution (R = 100000) integral field spectroscopy from 2.9 to 5.3~$\mu$m. Assuming a collecting aperture of 39\,m in diameter, METIS will have approximately six times more collecting power than the VLTI but five times poorer angular resolution. The two instruments will therefore probe complementary parameter spaces for a given wavelength.  Alternatively, METIS will provide an angular resolution in the near-infrared similar to that of Hi-5 in the thermal near-infrared (see Figure~\ref{fig1}). VLTI/Hi-5 will hence provide complementary high-contrast observations to characterize the observed planets and circumstellar disks. In particular, a VLTI instrument can make use of less-solicited telescopes such as the ATs to follow-up in the thermal near-infrared new ELT/METIS discoveries. Finally, note that, while METIS will obviously have the sensitivity to detect exozodis, it will hardly be possible to unambiguously detect extended sources within $\sim$3$\lambda$/D ($\sim$60\,mas at L band, $\sim$150\,mas at N band). Known hot and warm exozodis are all located much closer than that. Blackbody considerations also puts the dust closer than that for a G2V star located at 10\,pc ($\sim$15 mas for 800K dust better probed at L band and $\sim$100 mas for 300K dust better probed at N band). 

\item PFI (Planet Formation Imager\cite{Monnier:2016,Kraus:2016,Ireland:2016}) is currently a science-driven, international initiative to develop the roadmap for a future ground-based facility that will be optimised to image planet-forming disks on the spatial scale where the protoplanets are assembled, which is the Hill sphere of the forming planets. The goal of PFI will be to detect and characterise protoplanets during their first $\sim 100$ million years and trace how the planet population changes due to migration processes, unveiling the processes that determine the final architecture of exoplanetary systems. With $\sim 20$ telescope elements and baselines of $\sim 3$~km, the PFI concept is optimised for imaging complex scenes at thermal near-infrared and mid-infrared wavelengths (3-12$\mu$m) and at 0.1 milliarcsecond resolution. Hence, Hi-5's mission will be ``explorative", while PFI's mission will be to provide a comprehensive picture of planet formation and characterisation (resolving circumplanetary disks). Hi-5 and PFI will also share many common technology challenges, for instance on thermal near-infrared beam combination, accurate/robust fringe tracking, and nulling schemes. The proposed Hi-5 project will thus provide important data and needed information during the design and construction of PFI.

\end{itemize}

\section{Summary and conclusions}

The VLTI currently achieves contrasts of a few 10$^{-3}$ in the near-infrared and second-generation instruments are not designed to do better. Achieving deeper contrasts at small inner working angles is however mandatory to make scientific progress in various fields of astrophysics and, in particular, in exoplanet science. On the VLTI, gaining one order of magnitude (i.e., contrasts of at least a few 10$^{-4}$) is today within reach as demonstrated with ground-based nulling interferometers in the northern hemisphere and better contrasts (a few 10$^{-5}$) are theoretically possible with new promising ideas such as Kernel nulling. Besides the clear scientific motivation to reach such contrast levels, a new VLTI imaging instrument would be a very useful technology demonstrator for future major interferometric instruments such as PFI and TPF-I/DARWIN-like missions. 

\begin{acknowledgements}
The authors acknowledge the support from the H2020 OPTICON Joint Research Network. DD and OA thank the Belgian national funds for scientific research (FNRS). SK acknowledges support from an ERC Starting Grant (Grant Agreement No.\ 639889) and STFC Rutherford Fellowship (ST/J004030/1).
\end{acknowledgements}

\bibliographystyle{spiebib} % makes bibtex use spiebib.bst

\end{document}